\documentclass[aps,prl,twocolumn,groupedaddress]{revtex4-2}
\usepackage{graphicx}
\begin{document}

%\preprint{}

\title{Signatures of a Bilayer Structure in the Photoelectron Spectrum of $B_{80}^-$}

\author{Yi-Sha Chen}
\author{Jing-Jing Guo}
\author{Peng-Bo Liu}

\author{Hui-Yan Zhao}
\email{hyzhao@hebtu.edu.cn}
\thanks{Corresponding author}

\author{Jing Wang}

\author{Ying Liu}
\email{yliu@hebtu.edu.cn}
\thanks{Corresponding author}

\affiliation{Department of Physics and Hebei Advanced Thin Film Laboratory,
	Hebei Normal University, Shijiazhuang 050024, Hebei, China}


\date{\today}

\begin{abstract}
Recent photoelectron spectroscopy of B$_{80}^-$ was interpreted in terms of a fullerene-like cage structure. During our systematic investigation of medium-sized boron clusters, we identified a $D_{3h}$-symmetric bilayer isomer whose simulated photoelectron spectrum reproduces the principal features of the experimental photoelectron spectrum within 0.04 eV. The bilayer is energetically competitive with previously proposed structures and remains dynamically stable up to 1400 K according to $ab$ $initio$ molecular dynamics and vibrational analyses. Its electronic structure exhibits a 0.72 eV HOMO–LUMO gap and strong interlayer aromaticity, reflected by a NICS(0) value of $-44.3$ ppm in the interlayer B$\text{–}$B bonding region. These findings reveal a stable bilayer motif in the B$_{80}$ energy landscape and support its viability as a possible alternative structural assignment for the experimentally observed B$_{80}^{-}$.
\end{abstract}

%\keywords{}

\maketitle

\section{Introduction}

Boron, characterized by three valence electrons (2s$^2$2p$^1$), is intrinsically electron deficient and exhibits a unique combination of localized two-center two-electron (2c–2e) and delocalized multicenter two-electron ($m$c–2e) bonding. This unique bonding nature not only compensates for the electron shortage but also gives rise to a remarkable structural diversity in boron clusters. Extensive research has mapped out diverse topological configurations of boron nanostructures, including quasi-planar, tubular, cage, core-shell, and bilayer geometries \cite{ref1,ref2,ref3,ref4,ref5}. In 2007, Yakobson and co-workers proposed the hollow B$_{80}$ fullerene, stimulated extensive studies of boron nanostructures and related low-dimensional materials \cite{ref6,ref7,ref8,ref9,ref10,ref11,ref12,ref13,ref14,ref15,ref16,ref17,ref18,ref19}. Subsequent investigations identified a lower-energy core-shell structure incorporating an inner B$_{12}$ icosahedron\cite{ref20,ref21}, and similar motifs were later proposed in larger boron clusters, including B$_{92}$ \cite{ref22}, B$_{96}$ \cite{ref23}, B$_{98-102}$ \cite{ref24}, and B$_{111-114}$ \cite{ref25}.

Elucidating the precise geometries of boron clusters remains a formidable challenge due to their unconventional delocalized bonding. Joint photoelectron spectroscopy (PES) and first-principles calculations have established a size-dependent progression from planar/quasi-planar ($n = 3$--38, 41--42) to cage-like ($n = 39$--40) structural motifs \cite{ref3,ref26,ref27,ref28,ref29,ref30,ref31,ref32}, exemplified by the structural characterization and assignment of the quasi-planar B$_{36}^{-}$ cluster (a borophene precursor) \cite{ref33,ref34,ref35} and the cage-like B$_{40}^{-}$ borospherene \cite{ref26}. Recently, the structural paradigm expanded further with the discovery of the B$_{48}$ bilayer cluster, which features robust covalent interlayer bonds capable of extending into stable two-dimensional phases \cite{ref36,ref37}. This bilayer motif was further substantiated by the experimental synthesis of bilayer boron sheets on Ag and Cu surfaces \cite{ref38,ref39}. Concurrently, structural searches on medium-sized B$_n$ clusters ($n=52$--64) revealed a competition between bilayer and core-shell configurations, with the bilayer topology dominating at specific sizes such as $n=52$, 54, 60, and 62 \cite{ref40}. More recently, a highly stable B$_{63}$ bilayer configuration was proposed, emphasizing that configurations with fewer interlayer bonds (e.g., three versus four) can be energetically superior \cite{ref41}.

These collective developments have fueled a question: does the bilayer motif persist or even dominate in larger, historically significant clusters such as B$_{80}$? While various B$_{80}$ isomers, including core-shell and volleyball-like structures, have been extensively debated \cite{ref6,ref20,ref21,ref42}, the recent experimental PES characterization of B$_{80}^{-}$ by Choi and co-workers has reignited this exploration \cite{ref43}. Their study revealed a simple spectral pattern that was interpreted using a hollow buckminsterfullerene cage configuration \cite{ref43}. Intriguingly, during our systematic investigation of the structural evolution of medium-sized boron clusters, we discovered that a novel $D_{3h}$-symmetric B$_{80}$ bilayer structure carries distinct spectral signatures that closely match the newly measured PES profile. This bilayer configuration yields vertical detachment energies that reproduce the observed X, A, and B bands within 0.04 eV while maintaining structural integrity up to 1400 K. In this Letter, we present a comprehensive analysis of the stability, bonding characteristics, and electronic properties of this B$_{80}$ bilayer isomer, providing an alternative structural assignment for B$_{80}^{-}$ and offering new insights into topological competition and chemical bonding in large boron nanostructures.

\section{Results and Discussion}
\textit{Bilayer structure and stability.}---High-resolution geometric optimizations based on density functional theory (DFT) [see Supplemental Material \cite{ref44} for computational details] confirm that this novel atomic arrangement converges into a highly ordered $D_{3h}$ bilayer configuration. Vibrational frequency analysis reveals a complete absence of imaginary frequencies, with normal modes spanning from 141.1 to 1327.0 cm$^{-1}$, thereby confirming that this bilayer architecture corresponds to a true local minimum on the potential energy surface. The simulated Raman spectrum [Fig.~S1 in Ref.~\cite{ref44}] further highlights a prominent collective mode at 1003.0 cm$^{-1}$, which encapsulates the coherent synchronized vibration of the entire bilayer framework.

\begin{figure}[t]
	\centering
	\includegraphics[width=0.95\linewidth]{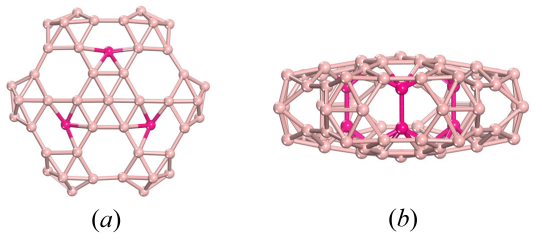}
	\caption{(Color online) Geometric structure of the $D_{3h}$-symmetric B$_{80}$ bilayer isomer showing the ($a$) top view and ($b$) side view.}
	\label{fig:fig1}
\end{figure}

It is noteworthy that, despite extensive structural searches on B$_{80}$ reported in the literature \cite{ref43}, this bilayer motif has not been reported previously, suggesting that additional low-energy configurations may still remain to be explored in the B$_{80}$ energy landscape. As depicted in Fig.~1, the optimized configuration comprises two symmetrically equivalent quasi-planar boron sheets interconnected through three interlayer B--B covalent bonds, forming a compact sandwiched bilayer architecture. The calculated average interlayer bond length is 1.715~\AA{}, in good agreement with those reported for experimentally synthesized bilayer-$\alpha$ borophene on Ag(111) (1.8~\AA{}), bilayer-$\alpha^{+}$ borophene B$_{22}$ (1.76~\AA{}), and freestanding $\nu_{1/12}$ bilayer borophene (1.73~\AA{}) \cite{ref38,ref39,ref45}. Furthermore, the interior of the B$_{80}$ bilayer is composed of pentagonal-pyramidal B$_6$ and hexagonal-pyramidal B$_7$ units, consistent with the structural Aufbau principle governing stable boron networks \cite{ref46,ref47}.

To precisely map the static energetic standing of this newly identified local minimum within the broader cluster landscape, we benchmarked its total energy against three historically prominent reference isomers: the buckyball ($T_h$) \cite{ref6}, core-shell ($C_1$) \cite{ref20}, and volleyball ($T_h$) \cite{ref42} configurations. Single-point energy evaluations spanning a diverse matrix of electronic structure methods systematically map their relative stability (see Table~S1 in Ref.~\cite{ref44}). Within the DFT framework, the semi-local PBE along with the hybrid PBE0 and TPSSh functionals (all incorporating dispersion corrections) consistently predict an identical energetic ranking: $\mathrm{Core\mbox{-}shell}(C_1)>\mathrm{Bilayer}(D_{3h})>\mathrm{Buckyball}(T_h)>\mathrm{Volleyball}(T_h)$. To definitively resolve potential functional dependencies and determine the energetic ordering, we further employed high-level correlated wave-function methods. Both SCS-MP2 and DLPNO-CCSD(T) calculations confirm that the core-shell ($C_1$) $>$ bilayer ($D_{3h}$) stability hierarchy remains invariant, with the buckyball and volleyball configurations emerging as energetically comparable yet less stable.

The thermodynamic stability of the B$_{80}$ bilayer at finite temperatures is further supported by \textit{ab initio} molecular dynamics (AIMD) simulations performed in the NVT ensemble. The $D_{3h}$ framework maintains its structural integrity up to 1400 K throughout 8 ps trajectories. This thermal stability exceeds that of the buckyball and volleyball isomers, which undergo structural degradation at approximately 1325 K, and approaches that of the lowest-energy core-shell configuration (1500 K)  (see Figs.~S2--S5 in Ref.~\cite{ref44}).

\textit{Spectroscopic fingerprints and structural assignment.}---To evaluate possible structural assignments for B$_{80}^{-}$, we benchmarked the simulated photoelectron spectra of the bilayer and competing isomers against the available experimental measurements (Fig.~2 and Fig.~S6 in Ref.~\cite{ref44}). As shown in Fig.~2($a$), the experimental PES of B$_{80}^{-}$ reported by Choi \textit{et al.} exhibits three well-resolved bands, labeled X, A, and B, with vertical detachment energies (VDEs) of 3.2, 4.0, and 4.8 eV, respectively \cite{ref43}. The adiabatic detachment energy (ADE), extracted from the onset of the X band, was determined to be 3.1 eV. The theoretical PES profiles of the B$_{80}^{-}$ buckyball and bilayer structures, shown in Figs.~2($b$) and 2($c$), respectively, were generated using Gaussian broadening with a full width at half maximum (FWHM) of 0.15 eV. Both structures reproduce the principal experimental spectral features, namely the X, A, and B bands. For the buckyball isomer, PBE0 calculations yield ADE/VDE values of 3.09/3.10 eV, in good agreement with the experimental X-band values of 3.1/3.2 eV. Likewise, the $D_{3h}$ bilayer structure gives corresponding ADE/VDE values of 3.19/3.24 eV, closely matching the experimental measurements. The calculated HOMO--LUMO gap of the bilayer isomer is 0.72 eV, comparable to the experimentally observed separation of 0.80 eV between the X and A bands. Consistent with this electronic structure, the simulated spectral features near 4.1 and 4.7 eV reproduce the higher-energy A (4.0 eV) and B (4.8 eV) bands observed experimentally [Fig.~2($c$)]. The close agreement between the simulated and experimental spectra provides spectroscopic signatures consistent with a bilayer structural motif in B$_{80}^{-}$, supporting its consideration as a possible alternative structural assignment.

\begin{figure}[t]
	\centering
	\includegraphics[width=0.85\linewidth]{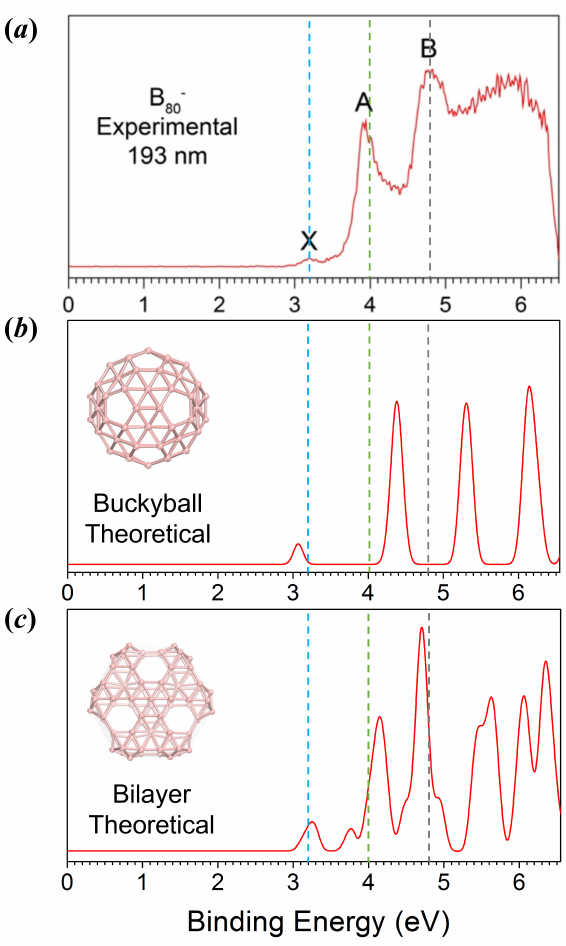}
	\caption{(Color online) ($a$) Photoelectron spectra (PES) of B$_{80}^-$. ($a$) Experimental spectrum recorded at 193 nm (6.424 eV), adapted from Ref.~\cite{ref43}. Simulated PES at the PBE0 level for ($b$) the buckyball and ($c$) the $D_{3h}$ bilayer isomer. The cyan, green, and gray dashed lines indicate the primary experimental peak positions (X, A, and B bands) at 3.2, 4.0, and 4.8 eV, respectively.}
	\label{fig:fig2}
\end{figure}

\begin{figure}[t]
	\centering
	\includegraphics[width=0.85\linewidth]{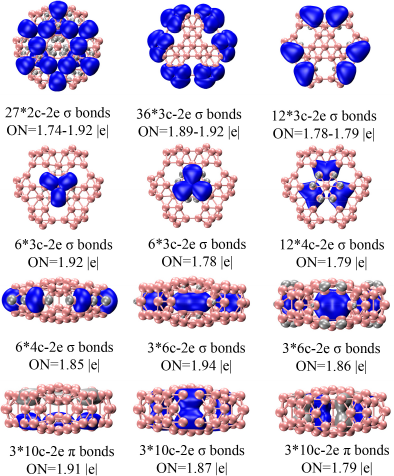}
	\caption{(Color online) AdNDP bonding analysis of the B$_{80}$ bilayer, ranging from localized peripheral 2c$\text{--}$2e bonds to highly delocalized 10c$\text{--}$2e bonding patterns. Corresponding occupation numbers (ONs) are indicated.}
	\label{fig:fig3}
\end{figure}

\textit{Electronic structure and chemical bonding.}---To uncover the quantum mechanical factors stabilizing the $D_{3h}$ bilayer architecture, we systematically examined the electronic properties and charge distributions of the neutral B$_{80}$ framework. Detailed mappings of the deformation electron density, partial density of states (PDOS), and frontier molecular orbitals are provided in the Supplemental Material [44]. The deformation electron density maps reveal a distinct charge depletion at the centers of the hexagonal rings, driving an outward charge transfer toward the peripheral boron framework, while strong electron localization anchors the interlayer bridging B-B bonds. Complementing this spatial picture, the HOMO-LUMO gap of 0.72 eV inherently reflects its chemical inertness. The PDOS demonstrates that the frontier molecular orbitals bounding this energy gap originate primarily from the hybridization of boron 2s and 2p atomic orbitals, with the 2p states exerting a dominant contribution near the Fermi level. To unearth the underlying multi-center bonding framework responsible for this exceptional electronic stability and high-temperature resilience, we next performed an adaptive natural density partitioning (AdNDP) analysis.

The resulting AdNDP framework partitions the 240 valence electrons into exactly 120 multi-center two-electron (mc-2e) $\sigma$ bonds, comprehensively mapping the cluster's chemical bonding landscape [Fig. 3]. Among the 27 localized 2c-2e $\sigma$ bonds, 3 are anchored along the interlayer B-B bridges, while the remaining 24 stabilize the hexagonal rings of both the upper and lower layers. Delocalized bonding is dominated by 3c-2e $\sigma$ interactions: 48 bonds (divided into two distinct types) brace the pentagonal pyramids at the twelve corners of the bilayer architecture, whereas another two types occupy the centered hexagonal B$_7$ units. Delocalized 12 4c-2e $\sigma$ bonds were located in the hexagonal pyramids and 6 delocalized 4c-2e $\sigma$ bonds were located in the B$_4$ rhombi at the waist. Two types of 6c-2e $\sigma$ bonds were situated in two different types of B$_6$ units at the waist. Furthermore, 9 10c-2e $\sigma$ bonds connected the upper and lower layers of the bilayer structure. This intricate network of localized and multi-center $\sigma$ bonds perfectly complements the spatial charge accumulation profiles resolved in the deformation electron density maps. To further manifest the magnetic signatures induced by these delocalized multi-center networks, we mapped the nucleus-independent chemical shift (NICS) profiles along two orthogonal pathways [Fig. 4], complemented by iso-chemical shielding surface (ICSS-ZZ) analyses [Figs. S10 and S11].

The NICS-scanned curves of the B$_{80}$ bilayer along the y- and z-axes confirm that the bilayer structure constitutes an aromatic system, with particularly strong aromaticity associated with the interlayer B--B bonds, which exhibit a NICS value of approximately $-44.3$ ppm in Fig.~4($a$). The NICS value in the central region of the upper and lower layers is about $-8.3$ ppm in Fig.~4($b$). The NICS scan along the y-axis shows two peaks at $\pm 5$~\AA{} with positive NICS values, corresponding to deshielded regions localized in the waist of the bilayer structure. These features are consistent with the ICSS$_{ZZ}$ results (iso-chemical shielding surface based on the calculated NICS-ZZ components), as illustrated in Figs.~S10 and S11. Additionally, the NICS scan along the z-axis exhibits symmetry, with a NICS value of approximately $-64.1$ ppm at the center of the hexagonal B units. The regions approximately 0.9~\AA{} above and below this central hexagon show positive NICS values, indicating antiaromatic character. The remaining regions with negative NICS values correspond to shielding areas, further confirming the aromatic nature of the bilayer structure.

\begin{figure}[t]
	\centering
	\includegraphics[width=0.95\linewidth]{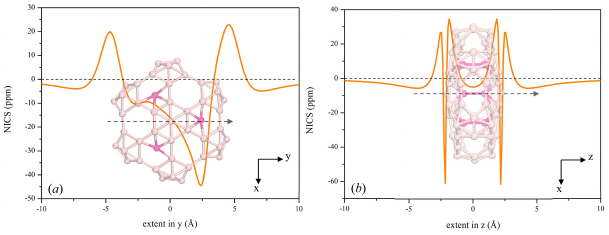}
	\caption{(Color online) NICS scan profiles along the ($a$) y axis and ($b$) z axis of the B$_{80}$ bilayer, illustrating the spatial evolution of its magnetic aromaticity.}
	\label{fig:fig4}
\end{figure}

\section{Conclusion}

In summary, we have systematically elucidated the structural, thermodynamic, and electronic properties of the $D_{3h}$ B$_{80}$ bilayer using first-principles calculations. The calculated average interlayer bond length closely matches that of bilayer borophene. Vibrational frequency analysis shows characteristic modes spanning 141.1 to 1327.0 cm$^{-1}$, and the absence of imaginary frequencies confirms that the structure corresponds to a true local minimum on the potential energy surface. \textit{Ab initio} molecular dynamics simulations further demonstrate that the B$_{80}$ bilayer remains dynamically stable up to 1400 K, exceeding the thermal stability of the buckyball and volleyball isomers ($\sim$1325 K) and approaching that of the core-shell configuration (1500 K). This stability is associated with global magnetic aromaticity, as evidenced by a diatropic shielding of approximately $-44.1$ ppm localized in the interlayer B–B bonding region. The simulated spectroscopic properties reproduce the principal features of the experimental photoelectron spectrum, providing spectroscopic evidence consistent with the possible presence of a bilayer motif in the experimental B$_{80}^{-}$ cluster ensemble. These results broaden the structural landscape of B$_{80}$ and provide insight into the relationship between boron clusters and borophene-based materials.

\textit{Acknowledgments}---This work is granted by the National Natural Science Foundation of China (Grant Nos. 12174084 and 12404338) and the Doctoral Research Start-up Foundation of Hebei Normal University (Grant Nos. L2023B08 and L2025B12).
%\end{acknowledgments}

\textit{Data availability}---The data that support the findings of this study are available from the corresponding author upon request.

\bibliography{B80.bib}

\end{document}